\documentclass[conference]{IEEEtran}
\IEEEoverridecommandlockouts
\usepackage{amsmath,amssymb,amsfonts}
\usepackage{algorithmic}
\usepackage{graphicx}
\usepackage[hyphens]{url}  
\usepackage{xcolor}
\usepackage{caption} 
\usepackage{booktabs} 
\usepackage[style=ieee,dashed=false]{biblatex}
\usepackage[hidelinks]{hyperref}
\usepackage[T2A,T1]{fontenc}
\usepackage[utf8]{inputenc}
\usepackage[russian,english]{babel}
\usepackage{csquotes}
\newcommand{\todo}[1]{}
\renewcommand{\todo}[1]{{\color{red} TODO: {#1}}}
\def\BibTeX{{\rm B\kern-.05em{\sc i\kern-.025em b}\kern-.08em
    T\kern-.1667em\lower.7ex\hbox{E}\kern-.125emX}}
\begin{document}

\title{Money Over Morals: A Business Analysis of Conti Ransomware\\
}

\author{\IEEEauthorblockN{ Ian W. Gray}
\IEEEauthorblockA{
\textit{New York University}\\
}
\and
\IEEEauthorblockN{ Jack Cable}
\IEEEauthorblockA{
\textit{Independent Researcher}\\
}
\and
\IEEEauthorblockN{Benjamin Brown}
\IEEEauthorblockA{
\textit{University of Michigan}\\
}
\and
\IEEEauthorblockN{ Vlad Cuiujuclu}
\IEEEauthorblockA{
\textit{Flashpoint}\\
}
\and
\IEEEauthorblockN{ Damon McCoy}
\IEEEauthorblockA{
\textit{New York University}\\
}
}


\maketitle

\begin{abstract}

Ransomware operations have evolved from relatively unsophisticated threat actors into highly coordinated cybercrime syndicates that regularly extort millions of dollars in a single attack. Despite dominating headlines and crippling businesses across the globe, there is relatively little in-depth research into the modern structure and economics of ransomware operations.

In this paper, we leverage leaked chat messages to provide an in-depth empirical analysis of Conti, one of the largest ransomware groups. By analyzing these chat messages, we construct a picture of Conti's operations as a highly-profitable business, from profit structures to employee recruitment and roles. We present novel methodologies to trace ransom payments, identifying over \$80 million in likely ransom payments to Conti and its predecessor -- over five times as much as in previous public datasets. As part of our work, we publish a dataset of 666 labeled Bitcoin addresses related to Conti and an additional 75 Bitcoin addresses of likely ransom payments. Future work can leverage this case study to more effectively trace -- and ultimately counteract -- ransomware activity.
\end{abstract}


\section{Introduction}
Ransomware is a type of malware that encrypts the files on a victim's computer, and charges an extortion fee for the decryption key. Ransomware attacks have significantly increased over the past years with the addition of more adversarial groups, new extortion tactics, and more targeted attacks. In 2021, ransomware payments exceeded \$600 million USD, according to cryptocurrency analysis firm Chainalysis \cite{team_as_2022}. 

This has resulted in the emergence of large-scale Ransomware as a Service (RaaS) operations that have streamlined segments of their campaigns by dividing the work across different roles and responsibilities. This often encompasses affiliate models, where a core team responsible for developing malware leases it to others to deploy and infect potential victims. However, there has been little academically peer-reviewed analysis of modern ransomware operations. This lack of insight into backend information on RaaS campaigns has left the security industry inferring for years, on an anecdotal basis, how these threats operate.

In this paper, we perform an analysis of leaked chat messages and cryptocurrency addresses associated with Conti. Based on a report from Chainalysis, Conti is one of the most prolific ransomware groups and has attacked thousands of organizations~\cite{team_as_2022}. Conti's victims include critical infrastructure entities such as hospitals and food providers~\cite{team_as_2022}. Despite setbacks to the Conti ransomware collective, including self-proclaimed shutdowns and re-branding, they continually ranked in the top three ransomware groups for number of victims and volume of ransoms in 2020 and 2021~\cite{noauthor_lockbit_nodate}. 

The chat data was leaked by a Ukrainian security researcher in February 2022 in response to the Russian invasion of Ukraine~\cite{cnn}. The leak included over 168,000 messages from Conti's internal chat logs. The chat logs contain information pertaining to the inner workings of the group, such as discussions of malware development and victim negotiations. These chats contain a wealth of data to aid in the understanding of Conti's inner operations, including  associates' Bitcoin wallet addresses, employee recruitment processes, and delineation of roles and responsibilities.

Our analysis drives insights that can be leveraged by law enforcement and policymakers to aid in counteracting ransomware. For instance, just two exchanges -- one unidentified exchange and Gemini -- are responsible for over 90\% of identified payments to Conti. Likewise, Conti exhibits poor operational security, with its associates sending a large amount of salary payments to exchanges like Gemini and Binance that enforce Know Your Customer (KYC) regulations. These centralized points provide opportunities to trace ransomware actors and seize funds.

In this paper, we make the following contributions: 

\textbf{Economic on-chain measurement}. We manually annotate all 666 Bitcoin addresses present in the leak according to their function (e.g. salary or reimbursement) which we will publicly publish. After annotating, we then use on-chain transaction data to provide an analysis of Conti's bottom line, including estimated gross revenue, operating cost, salary per role, cash-out techniques, and relation to other cybercrime activity (like dark web marketplaces). As part of this analysis, we develop a methodology to identify ransom payments based on common proceed splitting behavior, which we use to identify \$83.9 million in new likely payments.

\textbf{Qualitative business structure analysis}. The chat logs also contain qualitative information on different roles and responsibilities of Conti. Along with the Bitcoin address annotations, we identified the roles and responsibilities within the collective. We assessed team composition from the chats, as well as the primary users based upon interactions within the chat logs. We also provide an analysis of their employee recruitment process and  challenges managers faced with employees that did not know the illicit nature of their employer.

\section{Background}

In this section, we describe the functional roles of the archetypal Ransomware as a Service operation. These roles are segmented into specialized tasks that fulfill different parts of the ransomware attack chain \cite{hutchings_configuring_2017}. We explore how these roles execute the respective parts of the ransomware campaign, from malware delivery to cashing out \cite{thomas_framing_nodate, huang_tracking_2018}. 

Ransomware operations require individuals to build, test, maintain, and deliver the malware, as well as maintain victim communications during the ransom. Once a victim pays a ransom in cryptocurrency, the attacker launders the funds through a variety of exchanges and third party services.

Since the introduction of the first public ransomware leak site in 2019, approximately 80 ransomware groups have created public leak sites, where they threaten to post victim data if victims fail to meets the terms of the extortion \cite{noauthor_allied_nodate}. There are ransomware groups that do not maintain leak sites, and thus this number is not exhaustive. There is also overlap in these operations, including code re-use, and re-branding that typically occurs after a significant ransomware incident \cite{noauthor_ryuk_nodate, noauthor_ransomware_nodate}. 

At a macro-level, RaaS operations are generally divided between \textbf{Ransomware Operators} and \textbf{Ransomware Affiliates}. The operators are typically salaried workers that recruit new members, develop the malware, advertise and sell access to their ransomware, and maintain the victim payment portal and leak site post-compromise. Affiliates are typically commissioned workers that license the malware for a fee or a percentage of the ransom payment. Their role is to target and compromise new victims, deliver and execute the ransomware, and handle victim negotiations. Affiliates have also been associated with lateral movement, persistence, and data exfiltration in a victim's network \cite{team_human-operated_2020, noauthor_ransomware_nodate}. 

\textbf{Management:} RaaS operations can encompass hundreds of specialized workers. They have been likened to a gig economy for their on-demand services provided by their affiliate structure. Additionally, many phases of the attack chain are facilitated by human decision-making  \cite{noauthor_ransomware_nodate, team_human-operated_2020}. The managers are responsible for the human effort, which includes human resources, hiring, finances, and payroll. Managers may also have cross-departmental responsibilities, and support the other lines of effort listed below.     

\textbf{Development and Infrastructure:} Illicit economies are dependent upon administrative work and maintenance to ensure uninterrupted operations through development and infrastructure \cite{collier_cybercrime_nodate}. System administrators and software developers are salaried workers, essential to ensure uninterrupted RaaS operations. This may include acquiring or developing software, virtual machines, servers, proxies, antivirus (to test malware against), and a variety of other tools. These roles also offer IT support functions. 

\textbf{Access Operations:} Access brokers may sell access to affiliates, who use the access to escalate privileges and move laterally within a victims network. Initial access brokers monetize access to victim's networks. RaaS collectives may have their own access brokers, or they may outsource to third parties for access-as-a-service. Initial access brokers employ a variety of tactics, techniques, and procedures to gain access to victims networks, including spear-phishing key members of an organization, compromised credentials or remote desktop protocols (RDP), and exploiting vulnerabilities \cite{david_secret_2020, noauthor_initial_2022, noauthor_exposing_2022}. Access operations may employ a variety of tools to deploy malware, including Emotet, IcedID, Trickbot, and BazarLoader.  

\textbf{Negotiations:} Affiliates are typically responsible for managing negotiations post-compromise through an admin panel included with the ransomware. Large corporations may employ the use of ransomware negotiators, which deal directly with the ransomware affiliates to transfer cryptocurrencies through exchanges. RaaS operators manage the public leak site, where details of the victim are included if they fail to pay within a given time period. The operators also control the processing of ransomware payments.

\section{Data}
\label{data}

Our analysis in this study uses both leaked data, public blockchain data, and an annotated set of Bitcoin addresses from Crystal Blockchain, a commercial blockchain analysis platform~\cite{crystal}. Table~\ref{tab:datasets} provides a brief description of these data sources.
When using leaked data, there can arise both ethical and validity
concerns. In this section, we provide an overview of the
datasets, discuss how we validated the data,
and talk about the ethical framework of our study.

\subsection{Description}

\begin{table*}[t]
    \centering
\begin{tabular}{c c c c} 
\toprule

 Source & Information  &  Explanation\\
 \midrule
 Leaked Chats & timestamps, message, participants & Leaked Chat logs from Conti Jabber server\\
 Bitcoin Transactions & addresses, amount, timestamp & Public Bitcoin Blockchain Data\\
 Crystal Blockchain & annotated Bitcoin addresses & Platform to investigate Bitcoin addresses\\
 \bottomrule

\end{tabular}
    \caption{Summary of Datasets}
    \label{tab:datasets}
\end{table*}

On February 27, 2022, the Twitter account @ContiLeaks began tweeting links to an anonymous file sharing service that contained information related to the Conti Ransomware collective. In addition to malware source code and other internal files, the account shared three files of chat logs: two files containing messages from Conti's Jabber server and one file containing messages from Conti's Rocket.Chat server.

The dataset that we used for our analysis only contained text (i.e., no images). The leaked chats cover the period from July 2020 to February 2022. The portion of the dataset we analyzed did not contain any Personally Identifiable Information (PII). We created a set of regular expressions to extract Bitcoin addresses and confirmed that they were valid addresses. Table~\ref{tab:data_summary} provides a summary of the the datasets that we analyzed.

\begin{table}[t]
    \centering
\begin{tabular}{c c c c c} 
\toprule

 Source & Time Period & Posts  &  Users & Addresses\\
 \midrule
 Jabber & 2020-06-21 - 2022-02-25 & 168,624 & 463 & 665\\ 
 Rocket.Chat & 2020-08-31 - 2022-02-26 & 88,110 & 248 & 1\\
 \bottomrule

\end{tabular}
    \caption{Summary of Leaked Conti Chat Logs}
    \label{tab:data_summary}
\end{table}

\subsection{Validation}

The leaked datasets have been extensively validated by the security community, including the fact that gaps in the chat logs correlate with times when Conti was disrupted by law enforcement~\cite{krebs1}. In our analysis, Bitcoin addresses included in the leak are consistent with previously-known Conti Bitcoin addresses, such as those in the Ransomwhere dataset~\cite{ransomwhere}, with addresses in the leak having received funds from both Conti payment addresses and Ryuk (another ransomware strain operated by the same threat group~\cite{crowdstrike_ws}). Furthermore, we do not observe any internal inconsistencies in the dataset.

\subsection{Ethics}
We reason about potentially harms of our study through the lens of the Menlo report~\cite{2012dittrichmrafd}. We have two primary ethical questions. The first is a high-level question concerning whether the data being leaked should \textit{prima facie} prohibit all subsequent uses of it. For example, should a researcher be prohibited from analyzing the Facebook leaks in understanding their policies? We believe that the potential benefits of our study to society outweigh the minimal increased risks of harm. 

We observe that this data is already broadly available and the knowledge of its existence, its association with the Conti organization, and information, such as online handles and amount of Bitcoin transactions, have been publicly documented. Also, there is likely little if any Personally Identifiable Information (PII) in this leak and we did not find any during our analysis. This was a criminal service and the usernames are pseudonyms that are intentionally difficult to link to the actual persons. Thus, there is a minimal risk of us creating any new harm from our analysis. 

To further manage any remaining harms we institute several safeguards. We did not attempt to deanonymize anyone in these leaks as part of our study. Also, we do not use the publicly-known real names of any Conti employees or affiliates.
\section{Methodology}
\label{methodology}

\subsection{Database Annotation}

Jabber, the Extensible Messaging and Presence Protocol (XMPP), is a popular messaging application in the cybercrime underground. The open source instant messenger supports strong encryption, and independent federated servers that are located around the world \cite{noauthor_why_2017}. Well-established cybercrime forums, like Exploit, run their own Jabber servers. The Conti collective also operated their own Jabber server: q3mcco35auwcstmt[.]onion. 

 Similar to other online messengers, the Conti Leaks often included short text that by itself was absent of any substantive content. The large number of users (n = 463) within the chats are often overlapping, and span different parts of the operation. Additionally, Russian cybercriminals often use specialized slang, dubbed \foreignlanguage{russian}{Феня} (Fenya), that is purposefully difficult for a layperson to understand as it provides shorthand, obfuscation, and signals group membership \cite{roman-faithfull_russian_2022}. To better prepare the leaked messages for scientific analysis, we included a mixed method analysis that included quantitative and qualitative data analysis. 

Our primary objective in analyzing the data is to conduct an economic on-chain analysis of the cryptocurrency addresses observed in the dataset. We conducted a regular expression search within the chat messages to identify all mentions of Bitcoin addresses. In total, we identified 665 Bitcoin addresses in the Jabber dataset and 1 Bitcoin address in the Rocket.Chat dataset. As a result, we primarily focused on the Jabber dataset for this analysis.

In order to provide context when annotating addresses, we included 10 messages in a conversation before and 10 messages after each mention of a Bitcoin address. Using this approach, we were able to augment machine-translated text with manual translations for Russian slang, label the context of the Bitcoin address to inform the follow-on economic and business analysis, and ascribe roles to the Conti ransomware operators through the context of the chat messages. 

To better understand the context of the messages, including the Russian cybercrime slang, one of our annotators is a native Russian speaker and expert in the criminal underground. Three of the authors annotated the addresses, with one author annotating each address. We maintained a Russian slang dictionary that annotators could reference throughout our analysis.

When reviewing the Bitcoin addresses, we annotated the addresses according to the following labels: 

\noindent\textbf{Salary}: The address is associated with a request for salary or payment. Associates in the chat will often request from a manager that a salary be transferred to a wallet. 

\noindent\textbf{Reimbursement}: The address is associated with a request for reimbursement for a variety of services. Associates may directly or indirectly request through a manager that funds be transferred to a wallet for reimbursement of various tools. 

\noindent\textbf{Ransom Payment Address}: The address is used to receive payment from a Conti ransomware victim. 

\noindent\textbf{Claimed Ownership}: A member of the Conti collective claimed to own the address.  

\noindent\textbf{Services}: Any services that we can identify being directly mentioned by the Conti collective. 

\noindent\textbf{Victim Name}: The name of the victim who made the payment.\\

\textit{Inter-Annotator agreement:} To ensure that our annotations were consistent across researchers, we randomly sampled 100 posts containing cryptocurrency addresses and conducted a blind annotation with 3 raters. We then measured Inter-Annotator Agreement (IAA) by computing Fleiss’ Kappa for all 3 raters, which yielded a score of 0.73, indicating substantial agreement~\cite{fleiss}. 

\subsection{Economic On-Chain Measurement}
\label{econ}

We obtained Bitcoin addresses from the Conti leaks as well as the Ransomwhere dataset~\cite{ransomwhere}. Ransomwhere is a public, crowdsourced dataset of ransomware payment addresses, which we use to understand the blockchain techniques of Conti. We then performed on-chain blockchain analysis, detailed here, on these addresses.

To enrich our data, we fetched incoming and outgoing transaction data for all addresses from the blockchain.com API~\cite{noauthor_blockchaincom_nodate}. We then calculated dollar values for transactions by multiplying the amount of Bitcoin transacted by the closing Bitcoin to USD exchange rate the date the transaction was made from the CoinDesk API~\cite{coindesk}. While we cannot know the exact amount the ransomware actors sold the Bitcoin for, this serves as an approximation and is consistent with previous work~\cite{huang_tracking_2018,oosthoek_tale_2022}.

Additionally, to understand the types of wallets the addresses have interacted with, we utilized Crystal Blockchain~\cite{crystal}. Crystal Blockchain is a blockchain analytics tool that offers insight into the ownership of Bitcoin addresses based on a variety of public sources~\cite{crystal2}. We fetched the source and destination entities for all addresses in the dataset.

\begin{figure}[htp]
    \centering
    \includegraphics[width=8cm]{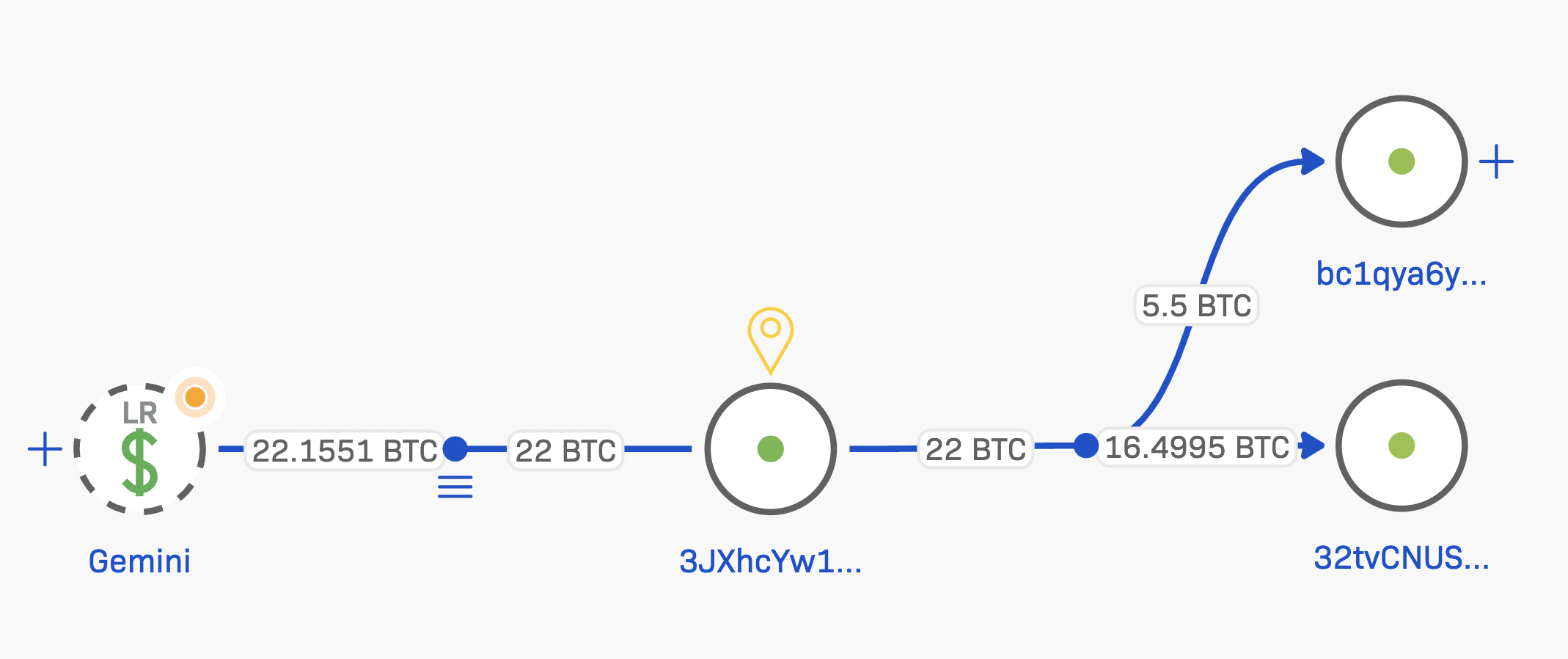}
    \caption{An example of splitting. This address received 22 Bitcoin from the US-based Gemini exchange, and split into 25\% and 75\%. 1 Bitcoin from this address would eventually be sent to an address in the leak. Other funds were transferred to other illicit entities, such as the sanctioned exchange Garantex.}
    \label{fig:splitting}
\end{figure}

In order to gain insight into the proceeds of Conti, we performed analysis to identify potential ransom payment addresses.  Based on confirmed Conti ransom payment addresses from Ransomwhere and those labeled in our dataset, we found 17 of 32 addresses to exhibit payment splitting, where the proceeds are immediately split to two wallets according to an exact percentage. This is likely due to the affiliate structure of Conti, where affiliates and the Conti core developers split proceeds. We found that for the 17 split addresses, split percentages ranged from 5\% to 40\%, with the most common (9 addresses) being 20\%. An example of a split payment is shown in Figure~\ref{fig:splitting}. Note that when we refer to split percentages, the percentage is the portion of the payment that the Conti collective keeps, with the remaining portion going to the affiliate.

In addition to low-risk exchanges such as Gemini, a large portion of these ransom payments to Conti originate from an unlabeled cluster of Bitcoin addresses. It is possible that this cluster belongs to an Over The Counter (OTC) desk, which many exchanges operate as a way for customers to exchange cryptocurrency outside of private markets. Given the significant portion of known Conti ransom payments originating from this cluster, it is possible that it is used by a common ransomware negotiator or incident response firm working with multiple victims. We consider this cluster in further analysis as a potential origin of Conti payments. Future work may attempt to identify the owner of this cluster.

We also analyzed 41 ransom payment addresses belonging to Ryuk from the Ransomwhere dataset. Ryuk is widely believed to be the predecessor to Conti, and both Conti and Ryuk have been attributed by Crowdstrike to be operated by the Wizard Spider group~\cite{ryuk_crowdstrike}. Of these 41 addresses, 17 exhibited splitting. Split percents ranged from 10\% to 50\%, with the most common (6 addresses) being 35\%.

To discover other likely ransom payment addresses, we considered addresses that: (1) sent money (directly or indirectly) to an address in the leaked dataset, (2) exhibited splitting according to an exact percent that was a multiple of 5 (e.g. 20\%, 25\%) and (3) had received more than 99\% of its funds from a low risk exchange (e.g. Gemini) or the identified unlabeled cluster. Results of this analysis are detailed in Section~\ref{economic_analysis}.

While we are able to designate these addresses as likely ransom payment addresses, the distinction between whether they are Conti or Ryuk is less clear. Through the course of the analysis, we observed previously known Ryuk addresses being used to fund addresses in the leaked Conti dataset, further suggesting that Conti and Ryuk are operated by the same actor. As Wizard Spider (an organized cybercrime group that has been attributed to Conti, Ryuk, TrickBot, and BazarLoader) paused operating Ryuk in March 2020, which coincided with the emergence of Conti, we label a likely ransom payment address as Ryuk if the address was first used before March 2020, and Conti otherwise~\cite{crowdstrike_ws}.

\subsection{Qualitative Business Analysis}

We extracted the unique aliases (463) from the Conti Leaks, and created a separate annotation document. We then read through the full dataset of the Conti Leaks (168,624 messages) and attempted to categorize the user roles based upon the content of their conversations. We found that a small number of individuals comprise a large number of the chats, so we sorted the aliases by degree centrality to understand who sent and received the most messages. We then decided to focus on the top 50 aliases, as most were also observed in our prior annotation of the cryptocurrency addresses. We maintained a full list of users, however we chose to focus annotations on the top 50. 

We made the following categories to understand their roles within the organizations: \textbf{Role}, \textbf{Direct Report}, \textbf{Working Relationships}, \textbf{Alternative Aliases}, and \textbf{Tasks}. While certain information regarding their respective roles could be gleaned from the chats, we had to otherwise infer based upon the context of the discussions, or their working relationships. 
\section{Economic Analysis}
\label{economic_analysis}
\label{results}



As with any business, Conti has income and expenses. The bigger the profit margin, the more its operators can walk away with. To begin our economic analysis, we utilize the labeled addresses to understand which addresses represent a business expense for Conti and which represent income. We consider reimbursements and salary to be expenses, while ransom payments are income.

\begin{table}[t]
    \centering
\begin{tabular}{c c c} 
\toprule

 Source & Amount &  Addresses\\
 \midrule
 Ransom payments in leaked dataset & \$3.4M & 5\\
 Ransom payments (Ransomwhere) & \$17.1M & 28\\
 Likely ransom payments (Conti) & \$57.4M & 41\\
 Likely ransom payments (Ryuk) & \$26.5M & 34\\
 \textbf{Total income} & \textbf{\$104.4M} & \textbf{107}\\
 \midrule
 Salary & \$21.9M & 419\\
 Reimbursement/Salary & \$5.4M & 15\\
 Reimbursement & \$3.8M & 227\\
 \textbf{Total expenses} & \textbf{\$31.2M} & \textbf{661}\\
 \bottomrule

\end{tabular}
    \caption{Conti income and expenses based on annotated Bitcoin addresses, Ransomwhere data, and inferred payments.}
    \label{tab:bitcoin_payments}
\end{table}

Table~\ref{tab:bitcoin_payments} shows the total income and expenses for Conti. Of the addresses in the leaked dataset, salaries represent the most in number (419) and the highest dollar value at \$21.9 million. Addresses that are used for both salary and reimbursements are relatively low in number but represent \$5.4 million in payments. Reimbursements, while lower in dollar value at \$3.8 million, have 227 addresses, suggesting that less money goes to reimbursement wallets on average than salary addresses.

Based on addresses in the leaks alone (the first row of Table~\ref{tab:bitcoin_payments}), expenses exceed income. This is to be expected, as Conti primarily used its administrator portal to communicate with victims, while the leaked chat logs appear to be the primary forum for requesting salary payment and reimbursement. As a result, ransom payment addresses surface in the chat logs only incidentally, while salaries and reimbursements are expected.

\begin{figure}[htp]
    \centering
    \includegraphics[width=8cm]{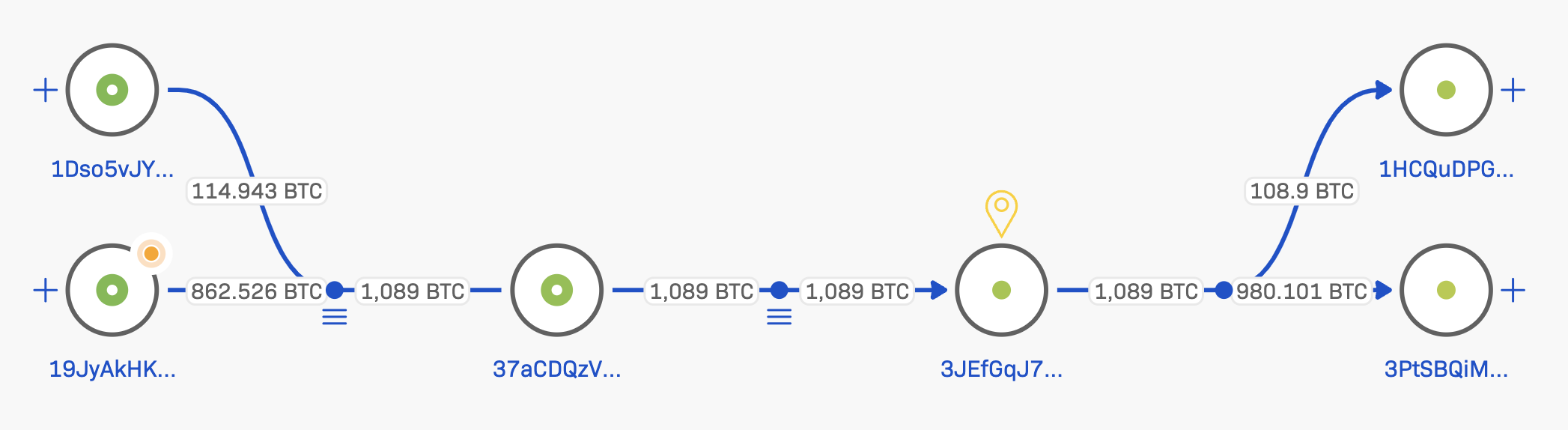}
    \caption{The largest discovered likely payment, of \$9.5M in March 2020. The funds originated from the unlabeled cluster discussed in Section~\ref{methodology}.}
    \label{fig:largest_payment}
\end{figure}

Nonetheless, we can identify likely ransom payment addresses. Given that Conti's income comes from ransom payments, and due to the traceable nature of Bitcoin, we can trace back payments visible in the leaked dataset to the ransom payments where the funds originate. Using the criteria established in Section~\ref{methodology}, we identify 75 likely ransom payment addresses representing \$83.9 million in payments. Of this, based on the dates Ryuk and Conti were active, we label \$26.5 million as Ryuk payments and \$57.4 million as Conti payments. The largest discovered likely payment of \$9.5M is shown in Figure~\ref{fig:largest_payment}.

Given this perspective, income begins to dwarf expenses. In addition to leftover money from Ryuk used to fund the Conti operation, the total Conti income (\$77.9 million) is more than double total expenses (\$31.2 million). A significant portion of the proceeds go directly to the hands of affiliates. Our numbers are likely incomplete -- Chainalysis identified \$180 million in proceeds from Conti in 2021 alone~\cite{team_as_2022}. However, unlike Chainalysis, we have provided our methodology for identifying ransom payment and we will publicly publish the addresses.

Table~\ref{tab:exchanges} shows the most common origins of confirmed and likely Conti ransom payments. The unlabeled cluster discussed in Section~\ref{methodology} represents a majority of payments -- almost 70\%. Following that, Gemini composes a significant share at \$23.1 million. The fact that just two exchanges represent the vast majority of identified payments to Conti suggests strong intervention points.

\begin{table}[t]
    \centering
\begin{tabular}{c c c c} 
\toprule

 Exchange & Confirmed Payments & Likely Payments & Total \\
 \midrule
 Unlabeled Cluster &  \$8.6M & \$64.8M & \$73.4M  \\
 Gemini & \$5.9M & \$17.4M & \$23.1M  \\
 Kraken & \$1.0M & \$0.2M & \$1.2M \\
 Coinbase & \$0.4M & \$0.6M & \$1.1M \\
 Binance & \$0.6M & \$0.0M & \$0.6M \\
 \bottomrule

\end{tabular}
    \caption{Top exchanges from which Conti ransom payments originate. Note that "Unlabeled Cluster" represents the unlabaled cluster of bitcoin addresses, discussed in Section~\ref{methodology}.}
    \label{tab:exchanges}
\end{table}

We have published the derived likely ransom payment addresses on GitHub.~\footnote{See \url{https://github.com/cablej/conti-payments}} Notably, the release of these addresses increases the amount of publicly known Conti payments more than fivefold -- from Ransomwhere's \$17.1 million to \$104.4 million.

\begin{figure*}[htp]
    \centering
    \includegraphics[width=\textwidth]{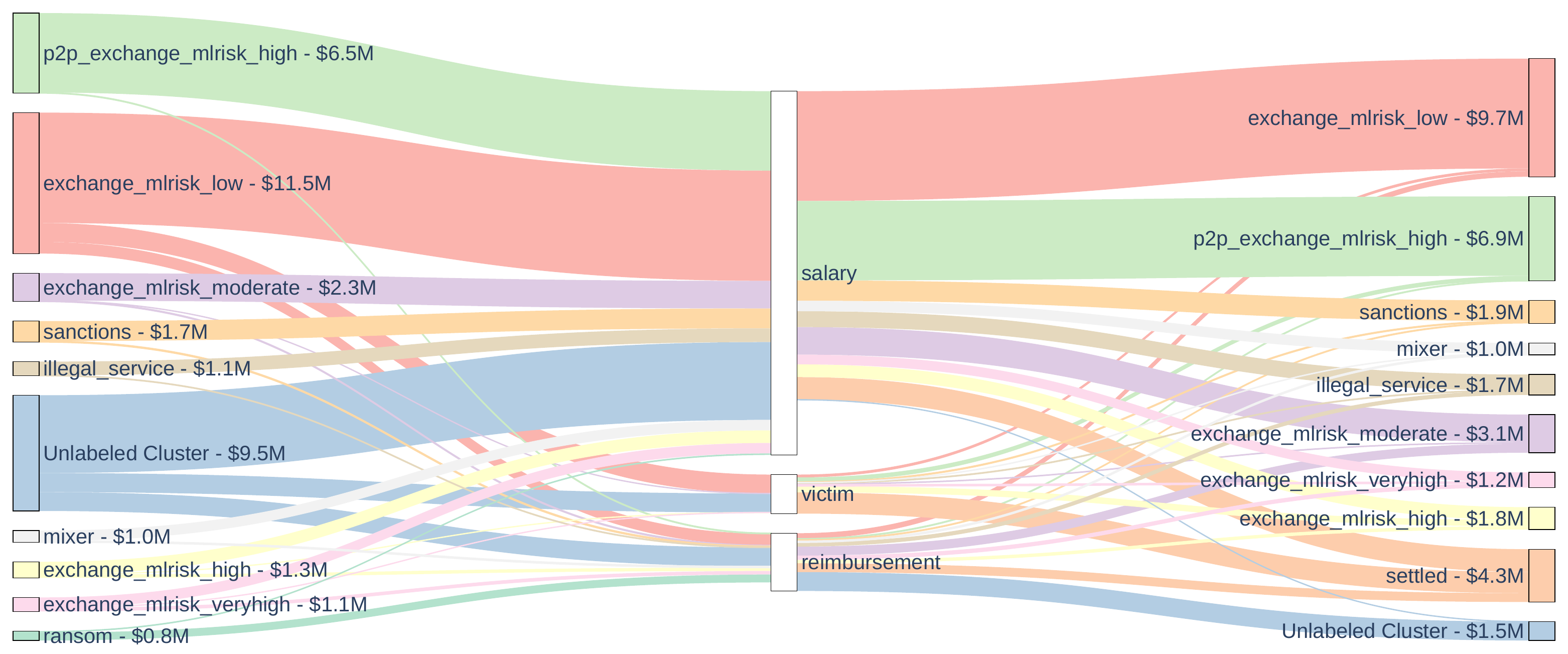}
    \caption{Labelled origins and destinations of wallet funds occurring in the Conti leaks dataset. Note that unknown addresses are excluded. ``mlrisk'' stands for money laundering risk. Further, note that as there are few ransom payment addresses in the Conti leaks dataset, the "victim" section in this chart only represents a fraction of all victim ransom payments to Conti.}
    \label{fig:sankey}
\end{figure*}

Next, we consider the sources and funds of funds from wallets in the leaked dataset, shown in Figure~\ref{fig:sankey}. We use money laundering risk levels provided by Crystal Blockchain to group exchanges into three categories of low, medium, and high risk. Consistent with the hypothesis that most money originates from victim payments, most money originates either from low risk exchanges or the unlabeled cluster. Moderate-high risk exchanges, sanctioned exchanges, illegal services, and mixers represent smaller amounts -- suggesting that some Conti actors might take steps to conceal their funds, though this practice is not systematized across the group. The receiving profile varies by type of address -- ransom payment funds come almost exclusively from low risk exchanges or the unlabeled cluster, while salaries and reimbursements represent a more diverse portfolio. We speculate that some salaries and reimbursements are paid from a slush fund belonging to the core operators, and thus have a variety of sources.

A large portion of wallet transactions, somewhat surprisingly, are to low risk exchanges -- exchanges most likely to abide by Know Your Customer (KYC) regulations. Gemini and Binance account for a large portion of these funds -- \$4.3 million and \$2.9 million, respectively. Given Gemini's position particularly as a regulated, U.S.-based exchange, Conti actors may have jeopardized their operational security by trading there. Other funds wind up in a variety of illicit destinations, such as \$6.8M in Ren Exchange, a peer-to-peer cross-blockchain exchange that can be used to launder funds, \$2.8M in the Seychelles-based exchange Huobi, and \$1.4M in the now-sanctioned Hydra marketplace.

Of the expanded set of ransom payments, the destination of funds includes a variety of services used to launder money. \$14.4M is sent to Ren Exchange, \$17.9M to Huobi, and \$12.6M to Binance. While both Huobi and Binance enforce KYC, certain illicit exchanges such as the now-sanctioned Suex have operated "nested exchanges" within both exchanges, providing an opportunity to launder funds through otherwise-regulated exchanges~\cite{suex}.

The leaks also offer insight into the individual salaries of Conti associates. Based on addresses where a Conti associate appeared to have claimed ownership of the address -- most often a salary address -- we compute the highest-grossing associates to be tramp (\$1.2M), mango (\$470K), baget (\$400K), bullet (\$280K), and andy (\$98K). We note that this is an incomplete view into the proceeds of these associates. 

We also observe evidence of co-spending among some associates. Co-spending occurs when two Bitcoin addresses are used as input to the same transaction, suggesting that the same entity controls both addresses. We observe two clusters of associates -- viper, jumbo, ganesh and sonar, and sticks, stakan, elvira, and bekeeper. It is likely that these two clusters use a shared Bitcoin wallet to manage their funds, or otherwise share ownership of funds.

\section{Business Analysis}

\subsection{Overlap and Re-branding}

Conti is assessed to be the successor of the Ryuk RaaS collective, which largely down-scaled their operations in March 2020 \cite{crowdstrike_ws}. This is evidenced from the leftover revenue that we identified that was likely used to fund Conti. Ryuk and Conti shared multiple features, most notably the use of Trickbot for initial infection. 

It is well documented that Trickbot and Conti are both technically and operationally interconnected \cite{noauthor_trm_nodate}. This overlap is significant to understand some of the roles and structures within Conti, because there are shared group members duties. Trickbot provided the initial infection and facilitated the installation of the Conti ransomware on a victim's machine, similar to Ryuk \cite{witte_presently_nodate}. An arrest warrant for a member of the Trickbot collective, max, indicates that a large number of Trickbot's members had also collaborate on the Dyre Trojan, a precursor to Trickbot. The remaining members of the Dyre collective transitioned to Trickbot following Dyre's takedown in 2015 \cite{witte_presently_nodate}. max's alias was identified within the Conti Leaks, also indicating overlap with Trickbot and Conti. Further, the Conti Leaks Twitter account leaked information from both Trickbot and Conti, including Trickbot's wider membership, indicating that there is approximately 18\% overlap with those Trickbot aliases within Conti's Jabber. 

The indictment of Trickbot malware developers max and follow-on indictment of ffx provided further details into the Trickbot organization, which also helped inform our understanding of Conti. Some of the same roles and responsibilities that were observed within the Trickbot organization were also observed within Conti, indicating that it was likely a rebranding as opposed to a reorganization. Trickbot and Conti also shared similarities in their roles, responsibilities, and recruiting methods \cite{witte_presently_nodate,noauthor_russian_2021}. 

\begin{figure}[htp]
    \centering
    \includegraphics[width=6cm]{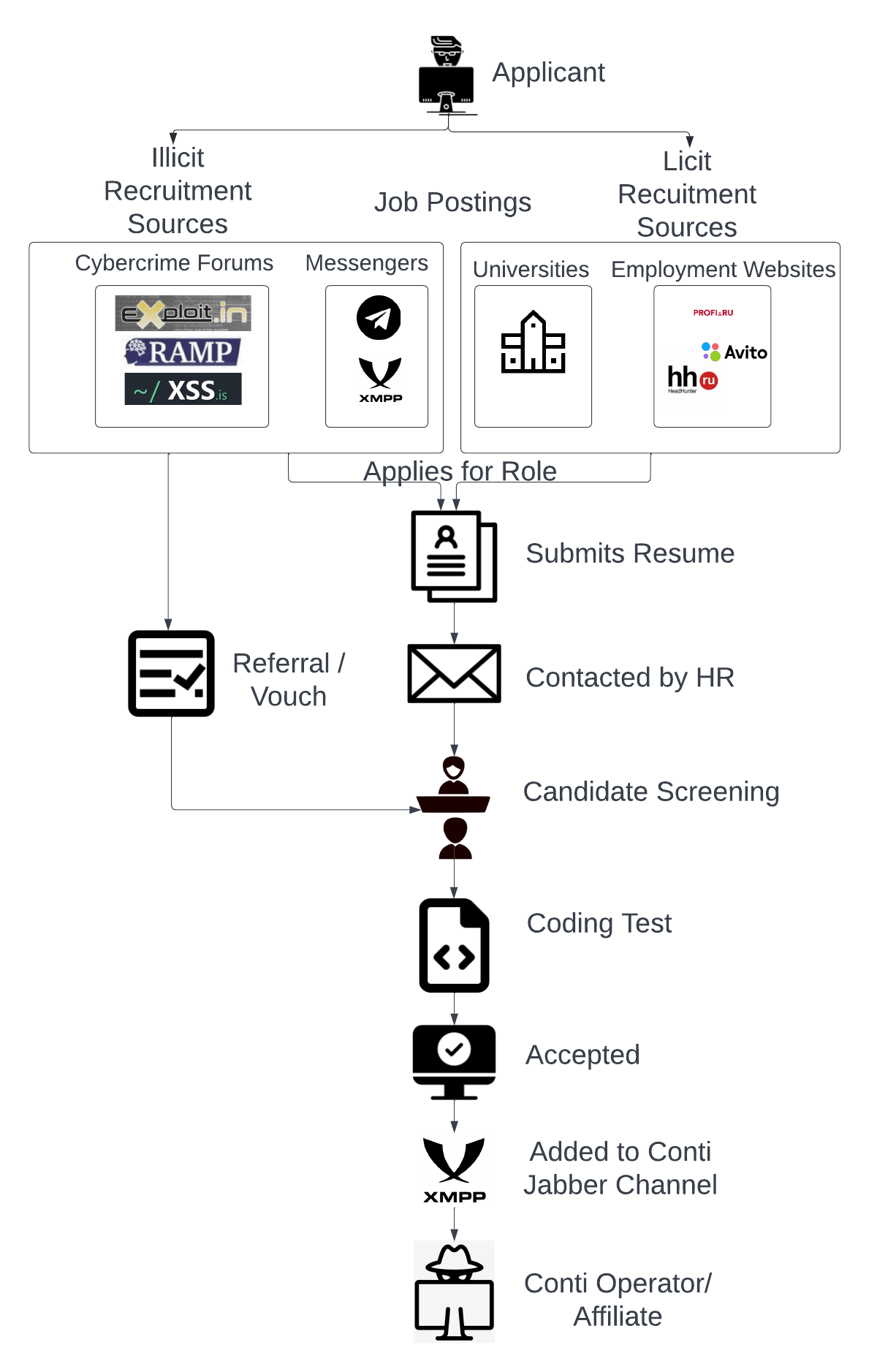}
    \caption{A flow chart demonstrating the recruitment sources of a RaaS affiliate }
    \label{fig:panel}
\end{figure}

\subsection{RaaS Roles, Responsibilities, and Recruiting}

Similar to RaaS collectives, Trickbot relied upon a network of specialized workers to facilitate different functions. For example, the unnamed defendants in the Trickbot indictment included the following roles: 

\begin{itemize}
  \item \textbf{Malware Manager:} Recruiting, hiring, testing malware, and procuring infrastructure
  \item \textbf{Malware Developer:} Oversaw functionality within the development of the malware
  \item \textbf{Crypters:} Encrypted the malware to prevent detection from anti-virus
  \item \textbf{Spammers:} Deployed the malware through targeted and broad-based phishing campaigns
\end{itemize}

According to the indictment, Trickbot advertised these roles on legitimate job posting websites, like LinkedIn and Indeed, as well as Russia-based freelance websites. After completion of a programming test, users were added to a private Jabber OTR communication server where they collaborated on "development, maintenance, and deployment of Trickbot." This is consistent with our observations of the recruiting methods used by Conti, which included recruiting for licit roles on job posting website like Avito, HeadHunter, and Profi[.]ru. Conti utilized similar recruiting methods, as observed in their Jabber, and select threads on underground forums.

On August 5, 2021, a disgruntled Conti affiliate m1Geelka leaked internal training materials, and IP addresses of their Cobalt Strike servers on XSS, a top tier underground forum. m1Geelka also commented on an IT recruitment thread from a user IT\_Work, stating that it was an advertisement to work with Conti. Between June 10, 2021 and September 6, 2021, IT\_Work had posted multiple offerings on underground Russian language forums, like XSS and Exploit, advertising seemingly legitimate job roles to support large IT projects. In our research, we assessed that these advertisements for licit roles were in concert with job postings on Russian-based freelance websites. 

\begin{itemize}
    \item \textbf{C++ Programmer} (with reverse engineering skills)
    \item \textbf{Full-stack web developer for PHP, NodeJS}
    \item \textbf{Windows System Administrator} 
    \item \textbf{Data Analyst} 
    \item \textbf{Business Analyst} 
    \item \textbf{UI/UX Designer} 
    \item \textbf{HTML Designer}
    \item \textbf{Pentester}
\end{itemize}

IT\_Work's posts demonstrate that while RaaS collectives are commonly associated with illicit tasks, like malware management and development, they also rely on technical talent to maintain infrastructure. These seemingly licit advertisements, albeit on underground forums, allowed Conti to recruit witting and unwitting tech workers to support the infrastructure of their operation. 

Following the Colonial Pipeline ransomware incident on May 6, 2021, President Biden threatened action against "ransomware networks \cite{house_remarks_2021}." As a result, XSS, Exploit, and Raid Forums banned ransomware advertisements. The leader of the former Babuk ransomware collective then started their own dedicated ransomware forum in May 2021, originally dubbed Payload.bin. The site changed its name to RAMP (Ransom Anon Market Place). While originally a closed forum composed of reputable threat actors, RAMP became public in August 2021 following an extortion attempt. Ransomware advertisements continued to be available on Telegram and Jabber \cite{team_after_2021}. 

Conti suffered a minor disruption in November 2021 after details of their infrastructure were reported on by a security firm \cite{malwarehunterteam_malwrhunterteam_conti_2021}. Shortly thereafter, a user JordanConti surfaced on RAMP highlighting that they were undeterred by the disruption, which included "peripheral IPs and wallets." JordanConti began openly recruiting for illicit roles required for their ransomware operation on RAMP, listing the Russian language as a requirement. The following roles were advertised on RAMP: 

\begin{itemize}
    \item \textbf{Pentesters:} "Top networkers who know how to bypass problematic AVs like Sentinel, work with RMM (Remote Monitoring Management) and EDR and backups"
    \item \textbf{Bot herders:} "Ideally, people with their own botnet, with a sufficient number of corp bots, especially in the US."
    \item \textbf{Targeted Spammers:} "who could beautifully send letters for "individual recipients" - the priority is USA."
\end{itemize}

From the Conti Leaks, we were able to ascertain that their HR specialists were also continuing to recruit on Russian-language freelance job posting websites and specialized universities. According to the Jabber chat logs, details of the roles varied. In a conversation between viper, a hiring representative, and bourbon, a developer, the reasoning varied from "we do pentesting for big clients," to more vague responses like "the work is remote, communication via messenger, the nature of the work is specific. That's all I know about conditions." viper then specified, "We do pen testing, write hacker software - exploits, grabbers, spam bots and more."

The legitimacy of the work was often questioned throughout the Conti leaks, as workers wondered why they had to be paid in cryptocurrencies, only communicated through encrypted messenger, and were unaware of the name or actual function of their employer. It does not appear that Conti used front companies to obscure their operations, but relied upon their managers to convey the appropriate messaging of the work. This meant deceiving their employees, or providing indirect answers to describe the true nature of their work.

\subsection{Team Composition}

From the chats, it appears that Conti is divided into several sub-teams. These teams are generally divided into functional areas, including management, development and infrastructure, access operations, and negotiations. These roles are consistent with the ransomware team structure outlined in the background.

In a chat from mango, a manager that oversees development and infrastructure, to stern, the organizational leader, mango shares details of the structure of his team, along with budgeted salaries:\\

\begin{quote}
the main team - \$97,447; 52 people\\
new team - \$4,000; 3 people, one has not started yet\\
reverse engineering - \$23,347; 16 people\\
research team - \$12,500; 6 people\\
osint intelligence team - \$9,000; 4 people\\
\end{quote}

mango's team does not appear to encompass the whole Conti operation, however one functional area. The total monthly salary for their team is assessed to be \$146,294.

Other references to a team structure appeared throughout the chat. For example, in a conversation between target and poll, target asks if poll needs individuals to attack logistics and manufacturing. poll highlights that they have a team that only "locks defense/military companies." In this regard, it appears that the sector specific targeting is divided between sub-divisions. However, some sector targeting like healthcare appeared to be off-limits.

Among RaaS operations as a whole, operators have informally agreed not to target healthcare. Following the DarkSide attack on Colonial Pipeline, REvil announced several new self-imposed restrictions for its operators and their affiliates. These announced restrictions included not targeting social sectors (such as healthcare and education) or any government entities, as well as requiring ransomware affiliates to get REvil operators’ approval for any future targets. In an interview, LockBit claimed that they have a, "negative attitude towards those who encrypt medical and educational institutions." In an exchange between reshaev, one of Conti's main developers, and  pin, who is possibly an affiliate, pin defends their reasoning for targeting a sports treatment center, claiming that it has no resuscitation unit and they have over 3K in insurance. reshaev emphasizes that they have a policy prohibiting ransoming healthcare, and recommends that pin "goes around them now." Despite this assertion, Conti had ransomed the healthcare sector through their operation including Ireland's Health Service Executive (HSE) and Department of Health (DoH), presumably choosing money over morals. 

In the Conti leaks, there are abstract references to specific teams. For example, mango introduces themselves as "support C, manager for general issues of the team \ trick \ locker, now I'm looking for access to work for the gang." buza, a team lead of coders, in an exchange with hof, a technical manager, makes abstract reference to "rocket" and "A," likely meaning Rocket.Chat and team "A" (one of three teams). The Rocket.Chat messages, though not included in our primary research, did include details of the team composition of the access operations. The user alter briefly described the structure and responsibilities of teams A, B, and C. alter did not mention the size of the groups, however there were 54 unique aliases in that server.\\ 

\begin{quote}
    The current composition is divided into groups, each group is assigned a team leader (one or two depending on the size of the group).\\
\end{quote}

\subsection{Primary Actors}

To further understand the main actors within the Conti leaks Jabber, we sorted the aliases by degree centrality. The top five individuals within the chats, defender, stern, buza, mango, and bentley, are Conti managers controlling payments, operations, developers, and malware builds. These managers also fulfilled HR functions, often sending bulk messages to users with comments, queries, and reminders to share cryptocurrency addresses for payments. Users that had a lower degree centrality were likely affiliates or developers. The limited number of chat messages made their roles much more difficult to identify. Managers like buza identified their developers by role in bulk messages that included requests to continue working on a bug tracker. 

Defender also sent bulk messages, not identifying recipients by role, requesting for alternative forms of communication. This likely indicates that the leaked Jabber was likely a centralized communication channel, and other communication channels may have been used for more specialized operations, like the Rocket.chat and Trickbot Forum, which included details on using the Trojan. 

The managers have power of the purse. The top five users by centrality are assessed to be some of the primary leadership, since their role also included communications with the channel. Requests for funds typically occurred in the Conti leaks Jabber, with team leads requesting salaries and reimbursement from managers on the behalf of the individuals on their teams. This information helped inform us on the hierarchy of the roles, relationship between aliases, and an understanding of the team structure. 

However, unlike previous cybercrime research that described the importance of cybercrime cultural capital within communities, the allure of experience and experimentation, it appears that RaaS operations center around mundane tasks of operating infrastructure \cite{collier_cybercrime_nodate, holt_digital_2019, leukfeldt_human_2019, goldsmith_seductions_2022}. The most important members of the Conti operation appear to be the managers overseeing the collective work, administering salaries, and approving expenses for reimbursement. 

\subsection{Rewards for Justice }

On May 6, 2022, the Department of State offered a \$10 million reward for information leading to the identification or location of the members of Conti collective as part of the Rewards for Justice program. On August 11, 2022, they requested specific information on five individuals: \\

\begin{itemize}
  \item dandis: manager, crypters 
  \item professor (aka alter): ransomware negotiator 
  \item reshaev: manager, ransomware builds 
  \item target: manager, access operations 
  \item tramp: manager, operations \\
\end{itemize}

From our research, we identified these individuals as being highly technical managers concerned with crypters, ransomware builds and development, access operations, and victim negotiations. Most of these aliases also appears within the Trickbot leaks, indicating that this may have overlap with the aforementioned Trickbot investigation. These individuals were most likely selected based upon the value that they provide the Conti collective in achieving a competitive advantage in the RaaS landscape \cite{noauthor_conti_nodate-8}. 

On February 9, 2023 (following the initial publication of this paper), the United States and United Kingdom sanctioned several members of the Trickbot collective for their role in cybercrime and ransomware operations \cite{noauthor_united_2023}. These individuals also appeared in the Conti Leaks, through the primary aliases shared in the sanction, or alternative aliases that helped us identify their membership in the collective. The following individuals were included in the sanction: \\

\begin{itemize}
  \item bentley (aka ben): senior manager  
  \item baget: developer
  \item globus: developer 
  \item tropa (aka kerasid): money laundering 
  \item iseldor: malicious injects 
  \item mushroom: manager 
  \item strix: administrator \\
\end{itemize}

These sanctions demonstrate a continued focus on cybercrime and ransomware operations. While the Rewards for Justice identified many of the lead members of Conti by alias, the sanctions listed the seven individuals by name. These measures underscore the importance of human capital in building and maintaining modern ransomware operations. 
\section{Related Work}

In order to conduct our analysis of the Conti ransomware operation, we use and extend methodologies from cyptocurrency tracking, leaked cybercrime data, and ransomware analysis.

\subsection{Cryptocurrency Tracking}

Prior work has shown that Bitcoin wallets and transactions are often linkable to the same entity using several heuristics~\cite{6113303, 10.1007/978-3-642-39884-1_4, 10.1007/978-3-642-39884-1_2,10.1145/2896384}. These Bitcoin tracing heuristics have been implemented into a number of commercial cryptocurrency forensic analysis tools which also use techniques to label the owner of account clusters, such as Chainalysis, TRM Labs, Elliptic, and Crystal Blockchain. We use Crystal Blockchain's cryptocurrency forensic tools to perform analysis of Bitcoin accounts that we identify in the leaked Conti chat data.

Huang et al. conducted a two year end-to-end measurement of ransomare operations, tracing bitcoin from acquisition to ransomware payment. In this analysis, known-victim payments were identified through seed addresses, and were clustered with previously unknown-victims. The authors identified ransomware revenue exceeding \$16 million USD, and infrastructure that was used to cashout illicit proceeds \cite{huang_tracking_2018}. Paquet{-}Clouston et al. identify \$13 million USD in ransomware payments between 2013 and 2017~\cite{paquet}.

\subsection{Ransomware as a Service}

Oosthoek et al. analyze over \$100 million in ransom payments through a crowd-sourced dataset of ransomware addresses~\cite{oosthoek_tale_2022}. The authors characterized the shift from commodity ransomware to RaaS. Along with increased profits came a growing sophistication evidenced by faster time to launder funds and increased operational security practices. We build on this work by conducting an in-depth analysis of a single ransomware groups, which allows us to map over five times the amount of payments to Conti, in addition to operating costs which were not previously analyzed.

Previous work has also documented the practice of Ransomware as a Service groups "splitting" payments between the ransomware group and affiliates. Cong et al. document DarkSide's split percentage, which varies based on the size of the ransom payment~\cite{crypto_cybercrimes}. Regarding Conti, Elliptic noted a 22.5\% split for several Conti ransom payment addresses~\cite{elliptic_conti}.

\subsection{Conti}

To date, relatively little academic work has analyzed the Conti leaks. Cong et al. investigate the cryptocurrency activities of several notable ransomware groups, including Conti~\cite{crypto_cybercrimes}. The authors compile data from a variety of sources, including public and proprietary data. As part of their work, the authors discuss Conti's activity at a high level, including analyzing Conti's posting of victim data on leak sites. Our work builds on this paper by performing in-depth analysis of Conti's economic and business practices, including extracting and analyzing 666 Bitcoin addresses, compared to the 239 addresses the authors extracted.

Other industry groups have analyzed primarily the business aspects of the Conti leaks, including ForeScout, Secureworks, and Check Point~\cite{forescout,secureworks,checkpoint}.
\section{Conclusion}
\label{conclusion}

Our study of Conti presents a vignette into the structure of a modern Ransomware as a Service group. This is the first comprehensive crypto-economic analysis of the Conti leaks, based on our annotation of cryptocurrency addresses present in the leaks, on-chain analysis of cryptocurrency payments, and qualitative business assessment based upon user conversations.

Through our analysis, we developed a methodology to identify ransom payments based on common splitting behavior. We use this methodology to identify \$83.9 million in new likely payments and can help to better inform ransomware-affiliated payments through exchanges. Identifying these payments may assist cryptocurrency exchanges in blocking these payments, putting additional pressure on ransomware operators.  

We find significant leverage points in both economic and business areas. The fact that a significant portion of funds both are received from and sent to low-risk exchanges presents an opportunity to monitor and seize funds. Further, targeting the organizational leadership responsible for the recruiting, hiring, training, and administering the various business units and infrastructure can also have an impact on their ability to function. The affiliate structure additionally provides opportunities to disrupt the more technical operators of the ransomware, thereby preventing affiliates' ability to lease the malware or receive operational support.

\section*{Acknowledgment}

We thank the anonymous reviewers for their insightful and constructive suggestions and feedback, and Crystal Blockchain for providing access to their platform. Funding for this
work was provided in part by National Science Foundation grants 1844753 and 2039693.

\printbibliography

@article{fleiss,
 ISSN = {0006341X, 15410420},
 URL = {http://www.jstor.org/stable/2529549},
 author = {Joseph L. Fleiss},
 journal = {Biometrics},
 number = {3},
 pages = {651--659},
 publisher = {[Wiley, International Biometric Society]},
 title = {Measuring Agreement between Two Judges on the Presence or Absence of a Trait},
 urldate = {2022-09-01},
 volume = {31},
 year = {1975}
}

@InProceedings{10.1007/978-3-642-39884-1_2,
author="Ron, Dorit and Shamir, Adi",
editor="Sadeghi, Ahmad-Reza",
title="Quantitative Analysis of the Full Bitcoin Transaction Graph",
booktitle="Financial Cryptography and Data Security",
year="2013",
publisher="Springer Berlin Heidelberg",
address="Berlin, Heidelberg",
pages="6--24",
}

@article{2012dittrichmrafd,
  author = {Dittrich, D and Kenneally, E},
  title = {The Menlo Report: Ethical Principles Guiding Information and Communication Technology Research},
  howpublished = {\url{https://catalog.caida.org/paper/2012_menlo_report_actual_formatted}},
  note = {Accessed: 2022-9-20}
  
}

@article{10.1145/2896384,
author = {Meiklejohn, Sarah and Pomarole, Marjori and Jordan, Grant and Levchenko, Kirill and McCoy, Damon and Voelker, Geoffrey M. and Savage, Stefan},
title = {A Fistful of Bitcoins: Characterizing Payments among Men with No Names},
year = {2016},
issue_date = {April 2016},
publisher = {Association for Computing Machinery},
address = {New York, NY, USA},
volume = {59},
number = {4},
issn = {0001-0782},
url = {https://doi.org/10.1145/2896384},
doi = {10.1145/2896384},
journal = {Commun. ACM},
month = {mar},
pages = {86–93},
numpages = {8}
}

@InProceedings{10.1007/978-3-642-39884-1_4,
author="Androulaki, Elli
and Karame, Ghassan O.
and Roeschlin, Marc
and Scherer, Tobias
and Capkun, Srdjan",
editor="Sadeghi, Ahmad-Reza",
title="Evaluating User Privacy in Bitcoin",
booktitle="Financial Cryptography and Data Security",
year="2013",
publisher="Springer Berlin Heidelberg",
address="Berlin, Heidelberg",
pages="34--51",
isbn="978-3-642-39884-1"
}

@INPROCEEDINGS{6113303,
  author={Reid, Fergal and Harrigan, Martin},
  booktitle={2011 IEEE Third International Conference on Privacy, Security, Risk and Trust and 2011 IEEE Third International Conference on Social Computing}, 
  title={An Analysis of Anonymity in the Bitcoin System}, 
  year={2011},
  volume={},
  number={},
  pages={1318-1326},
  doi={10.1109/PASSAT/SocialCom.2011.79}}

@article{paquet,
  author    = {Masarah Paquet{-}Clouston and
               Bernhard Haslhofer and
               Benoit Dupont},
  title     = {Ransomware Payments in the Bitcoin Ecosystem},
  journal   = {CoRR},
  volume    = {abs/1804.04080},
  year      = {2018},
  url       = {http://arxiv.org/abs/1804.04080},
  eprinttype = {arXiv},
  eprint    = {1804.04080},
  bibsource = {dblp computer science bibliography, https://dblp.org}
}

@article{crystal2, url={https://crystalblockchain.com/frequently-asked-questions/}, title={Frequently Asked Questions}, author={{Crystal Blockchain}}}

@article{crystal, url={https://crystalblockchain.com/}, title="{Crystal} {Blockchain}"}

@article{ryuk_crowdstrike, url={https://adversary.crowdstrike.com/en-US/adversary/wizard-spider/}, title={Wizard Spider}, author="CrowdStrike"}

@article{coindesk, url={https://api.coindesk.com/v1/bpi/historical/close.json}, title={CoinDesk {API}}, author="CoinDesk"}

@article{crowdstrike_ws, url={https://www.crowdstrike.com/blog/wizard-spider-adversary-update/}, author="CrowdStrike", year=2020,month=10,title={{WIZARD SPIDER Update: Resilient, Reactive and Resolute}}}

@dataset{ransomwhere,
  author       = {Cable, Jack},
  title        = {{Ransomwhere: A Crowdsourced Ransomware Payment 
                   Dataset}},
  month        = may,
  year         = 2022,
  publisher    = {Zenodo},
  version      = {1.0.1},
  doi          = {10.5281/zenodo.6562484},
  url          = {https://doi.org/10.5281/zenodo.6562484}
}

@article{crypto_cybercrimes,
author="Cong, Lin and Harvey, Campbell R. and Rabetti, Daniel and Wu, Zong-Yu",
title="An Anatomy of Crypto-Enabled Cybercrimes",
year="2022",
month=July,
url={https://ssrn.com/abstract=4188661}
}

@article{elliptic_conti, url={https://www.elliptic.co/blog/conti-ransomware-nets-at-least-25.5-million-in-four-months}, title={{Conti Ransomware Nets at Least \$25.5 Million in Four Months}}, author={Elliptic}, year=2021,month=11}

@article{suex, url={https://www.coindesk.com/business/2021/10/05/heres-what-we-know-about-suex-the-first-crypto-firm-sanctioned-by-us/}, title={Here’s What We Know About Suex, the First Crypto Firm Sanctioned by US}, author={Baydakova, Anna}, year=2021,month=10}

@article{cnn, url={https://www.cnn.com/2022/03/30/politics/ukraine-hack-russian-ransomware-gang/index.html}, author="Lyngaas, Sean", year=2022, month="March", title="{I can fight with a keyboard': How one Ukrainian IT specialist exposed a notorious Russian ransomware gang}"}

@article{krebs1, url={https://krebsonsecurity.com/2022/03/conti-ransomware-group-diaries-part-i-evasion/}, title="Conti Ransomware Group Diaries, Part I: Evasion", author="Krebs, Brian", year=2022, month=3}

@article{forescout, url={https://www.forescout.com/resources/analysis-of-conti-leaks/}, title={Analysis of Conti Leaks}, year=2022, month=3, author="Forescout"}

@article{checkpoint, url={https://research.checkpoint.com/2022/leaks-of-conti-ransomware-group-paint-picture-of-a-surprisingly-normal-tech-start-up-sort-of/}, title={{Leaks of Conti Ransomware Group Paint Picture of a Surprisingly Normal Tech Start-Up… Sort Of}}, author={{Check Point}}, year=2022, month=3}

@article{secureworks, url={https://www.secureworks.com/blog/gold-ulrick-leaks-reveal-organizational-structure-and-relationships}, author={{Secure Works}}, title={{GOLD ULRICK Leaks Reveal Organizational Structure and Relationships}}, year=2022,month=3}

@misc{team_as_2022,
	title = {As {Ransomware} {Payments} {Continue} to {Grow}, {So} {Too} {Does} {Ransomware}’s {Role} in {Geopolitical} {Conflict}},
	url = {https://blog.chainalysis.com/reports/2022-crypto-crime-report-preview-ransomware/},
	language = {en-US},
	urldate = {2022-09-18},
	journal = {Chainalysis},
	author = {Chainalysis},
	month = feb,
	year = {2022},
}

@misc{noauthor_lockbit_nodate,
	title = {{LockBit}, {Conti}, and {BlackCat} {Lead} {Pack} {Amid} {Rise} in {Active} {RaaS} and {Extortion} {Groups}: {Ransomware} in {Q1} 2022 - {Security} {News}},
	shorttitle = {{LockBit}, {Conti}, and {BlackCat} {Lead} {Pack} {Amid} {Rise} in {Active} {RaaS} and {Extortion} {Groups}},
	url = {https://www.trendmicro.com/vinfo/us/security/news/ransomware-by-the-numbers/lockbit-conti-and-blackcat-lead-pack-amid-rise-in-active-raas-and-extortion-groups-ransomware-in-q1-2022},
 author = {{Trend Micro}},
	language = {en},
	urldate = {2022-09-23},
}

@inproceedings{hutchings_configuring_2017,
	title = {Configuring {Zeus}: {A} case study of online crime target selection and knowledge transmission},
	shorttitle = {Configuring {Zeus}},
	doi = {10.1109/ECRIME.2017.7945052},
	booktitle = {2017 {APWG} {Symposium} on {Electronic} {Crime} {Research} ({eCrime})},
	author = {Hutchings, Alice and Clayton, Richard},
	month = apr,
	year = {2017},
	note = {ISSN: 2159-1245},
	pages = {33--40},
}

@article{thomas_framing_nodate,
	title = {Framing {Dependencies} {Introduced} by {Underground} {Commoditization}},
	language = {en},
	author = {Thomas, Kurt and Huang, Danny Yuxing and Wang, David and Bursztein, Elie and Grier, Chris and Holt, Thomas J and Kruegel, Christopher and McCoy, Damon and Savage, Stefan and Vigna, Giovanni},
	pages = {24},
}

@inproceedings{huang_tracking_2018,
	address = {San Francisco, CA},
	title = {Tracking {Ransomware} {End}-to-end},
	isbn = {978-1-5386-4353-2},
	url = {https://ieeexplore.ieee.org/document/8418627/},
	doi = {10.1109/SP.2018.00047},
	language = {en},
	urldate = {2022-07-12},
	booktitle = {2018 {IEEE} {Symposium} on {Security} and {Privacy} ({SP})},
	publisher = {IEEE},
	author = {Huang, Danny Yuxing and Aliapoulios, Maxwell Matthaios and Li, Vector Guo and Invernizzi, Luca and Bursztein, Elie and McRoberts, Kylie and Levin, Jonathan and Levchenko, Kirill and Snoeren, Alex C. and McCoy, Damon},
	month = may,
	year = {2018},
	pages = {618--631},
}

@misc{noauthor_allied_nodate,
	title = {Allied {Universal} {Breached} by {Maze} {Ransomware}, {Stolen} {Data} {Leaked}},
	url = {https://www.bleepingcomputer.com/news/security/allied-universal-breached-by-maze-ransomware-stolen-data-leaked/},
 author="Abrams, Lawrence",
 year="2019", month="November",
	language = {en-us},
	urldate = {2022-08-24},
	journal = {BleepingComputer},
}

@misc{noauthor_ryuk_nodate,
	title = {Ryuk successor {Conti} {Ransomware} releases data leak site},
 author="Abrams, Lawrence",
	url = {https://www.bleepingcomputer.com/news/security/ryuk-successor-conti-ransomware-releases-data-leak-site/},
 year="2020",
 month="August",
	language = {en-us},
	urldate = {2022-08-28},
	journal = {BleepingComputer},
}

@misc{noauthor_ransomware_nodate,
	title = {Ransomware as a {Service} ({RaaS}) {Explained} {\textbar} {CrowdStrike}},
	url = {https://www.crowdstrike.com/cybersecurity-101/ransomware/ransomware-as-a-service-raas/},
	language = {en},
	urldate = {2022-09-22},
 year=2022,
 month=2,
 author="Baker, Kurt",
	journal = {crowdstrike.com},
}

@misc{team_human-operated_2020,
	title = {Human-operated ransomware attacks: {A} preventable disaster},
	shorttitle = {Human-operated ransomware attacks},
	url = {https://www.microsoft.com/security/blog/2020/03/05/human-operated-ransomware-attacks-a-preventable-disaster/},
	language = {en-US},
	urldate = {2022-09-21},
	journal = {Microsoft Security Blog},
	author = {Microsoft},
	month = mar,
	year = {2020},
}

@article{collier_cybercrime_nodate,
	title = {Cybercrime is (often) boring: maintaining the infrastructure of cybercrime economies},
	language = {en},
	author = {Collier, Ben and Clayton, Richard and Hutchings, Alice and Thomas, Daniel R},
	pages = {25},
}

@misc{david_secret_2020,
	title = {The {Secret} {Life} of an {Initial} {Access} {Broker}},
	url = {https://kela.local/the-secret-life-of-an-initial-access-broker/},
	language = {en-US},
	urldate = {2022-09-05},
	journal = {Kela},
	author = {David, Efrat},
	month = aug,
	year = {2020},
}

@misc{noauthor_initial_2022,
	title = {Initial access broker repurposing techniques in targeted attacks against {Ukraine}},
	url = {https://blog.google/threat-analysis-group/initial-access-broker-repurposing-techniques-in-targeted-attacks-against-ukraine/},
	language = {en-us},
	urldate = {2022-09-08},
 author = "Bureau, Pierre-Marc",
	journal = {Google},
	month = sep,
	year = {2022},
}

@misc{noauthor_exposing_2022,
	title = {Exposing initial access broker with ties to {Conti}},
 author = {Stolyarov, Vlad and Sevens, Benoit},
	url = {https://blog.google/threat-analysis-group/exposing-initial-access-broker-ties-conti/},
	language = {en-us},
	urldate = {2022-08-19},
	journal = {Google},
	month = mar,
	year = {2022},
}

@misc{noauthor_why_2017,
	title = {Why {Jabber} reigns across the {Russian} cybercrime underground},
	url = {https://www.cyberscoop.com/jabber-xmpp-cybercrime-russia-encrypted-chat/},
 author = {Howell O'Neill, Patrick},
 year=2017,month=4,
	language = {en},
	urldate = {2022-08-12},
	journal = {CyberScoop},
	month = apr,
	year = {2017},
}

@misc{roman-faithfull_russian_2022,
	title = {Russian prison culture and slang on cybercriminal forums: {Can} you cram on the hairdryer? {\textbar} {Digital} {Shadows}},
	shorttitle = {Russian prison culture and slang on cybercriminal forums},
	url = {https://www.digitalshadows.com/blog-and-research/russian-prison-culture-and-slang-on-cybercriminal-forums-can-you-cram-on-the-hairdryer/},
	language = {en-US},
	urldate = {2022-08-12},
	author = {Faithfull, Roman},
	month = may,
	year = {2022},
}

@misc{noauthor_blockchaincom_nodate,
author={Blockchain.com},
	title = {{Blockchain Data API}},
	url = {https://www.blockchain.com/api/blockchain_api},
	language = {en},
	urldate = {2022-09-07},
}

@misc{oosthoek_tale_2022,
	title = {A {Tale} of {Two} {Markets}: {Investigating} the {Ransomware} {Payments} {Economy}},
	shorttitle = {A {Tale} of {Two} {Markets}},
	url = {http://arxiv.org/abs/2205.05028},
	language = {en},
	urldate = {2022-07-12},
	publisher = {arXiv},
	author = {Oosthoek, Kris and Cable, Jack and Smaragdakis, Georgios},
	month = may,
	year = {2022},
	note = {arXiv:2205.05028 [cs]},
}

@misc{noauthor_trm_nodate,
	title = {{TRM} {Analysis} {Corroborates} {Suspected} {Ties} {Between} {Conti} and {Ryuk} {Ransomware} {Groups} and {Wizard} {Spider} {\textbar} {TRM} {Insights}},
	url = {https://www.trmlabs.com/post/analysis-corroborates-suspected-ties-between-conti-and-ryuk-ransomware-groups-and-wizard-spider},
 author={{TRM Labs}},
 month=4,year=2022,
	language = {en},
	urldate = {2022-08-28},
}

@misc{witte_presently_nodate,
	title = {Latvian {National} {Charged} for {Alleged} {Role} in {Transnational} {Cybercrime} {Organization}},
	url = {https://www.justice.gov/opa/pr/latvian-national-charged-alleged-role-transnational-cybercrime-organization},
	language = {en},
	urldate = {2022-09-23},
	month = jun,
	year = {2021},
}

@misc{noauthor_russian_2021,
	title = {Russian {National} {Extradited} to {United} {States} to {Face} {Charges} for {Alleged} {Role} in {Cybercriminal} {Organization}},
	url = {https://www.justice.gov/opa/pr/russian-national-extradited-united-states-face-charges-alleged-role-cybercriminal},
 author={{U.S. Department of Justice}},
	language = {en},
	urldate = {2022-08-23},
	month = oct,
	year = {2021},
}

@misc{house_remarks_2021,
	title = {Remarks by {President} {Biden} on the {Colonial} {Pipeline} {Incident}},
	url = {https://www.whitehouse.gov/briefing-room/speeches-remarks/2021/05/13/remarks-by-president-biden-on-the-colonial-pipeline-incident/},
	language = {en-US},
	urldate = {2022-08-30},
	journal = {The White House},
	author = {{The White House}},
	month = may,
	year = {2021},
}

@misc{team_after_2021,
	title = {After {Ransomware} {Ads} {Are} {Banned} {On} {Cybercrime} {Forums}, {Alternative} {Platforms} {Being} {Used} to {Advertise} and {Recruit}},
	url = {https://flashpoint.io/blog/avoslocker-ransomware-advertise-and-recruit/},
	urldate = {2022-08-30},
	journal = {Flashpoint},
	author = {Flashpoint},
	month = jul,
	year = {2021},
}

@misc{malwarehunterteam_malwrhunterteam_conti_2021,
	type = {Tweet},
	title = {MalwareHunterTeam on {Twitter}},
	url = {https://twitter.com/malwrhunterteam/status/1461450607311605766},
	language = {en},
	urldate = {2022-08-30},
	journal = {Twitter},
	author = {{MalwareHunterTeam [@malwrhunterteam]}},
	month = nov,
	year = {2021},
}

@article{holt_digital_2019,
	title = {Digital drift and the “sense of injustice”: counter-productive policing of youth cybercrime},
	copyright = {© 2018 Taylor \& Francis Group, LLC.},
	issn = {0163-9625},
	shorttitle = {Digital drift and the “sense of injustice”},
	url = {https://digital.library.adelaide.edu.au/dspace/handle/2440/119431},
	language = {en},
	urldate = {2022-09-23},
	author = {Holt, T. J. and Brewer, R. and Goldsmith, A.},
	year = {2019},
	note = {Accepted: 2019-06-11T01:47:37Z
ISBN: 9780030114939
Publisher: Taylor \& Francis},
}

@book{leukfeldt_human_2019,
	title = {The {Human} {Factor} of {Cybercrime}},
	url = {https://library.oapen.org/handle/20.500.12657/23615},
	language = {English},
	urldate = {2022-09-23},
	publisher = {Taylor \& Francis},
	editor = {Leukfeldt, Rutger and Holt, Thomas J.},
	year = {2019},
	note = {Accepted: 2019-12-09 13:48:21},
}

@article{goldsmith_seductions_2022,
	title = {The seductions of cybercrime: {Adolescence} and the thrills of digital transgression},
	volume = {19},
	issn = {1477-3708},
	shorttitle = {The seductions of cybercrime},
	url = {https://doi.org/10.1177/1477370819887305},
	doi = {10.1177/1477370819887305},
	language = {en},
	number = {1},
	urldate = {2022-09-23},
	journal = {European Journal of Criminology},
	author = {Goldsmith, Andrew and Wall, David S.},
	month = jan,
	year = {2022},
	note = {Publisher: SAGE Publications},
	pages = {98--117},
}

@misc{noauthor_united_2023,
	title = {United {States} and {United} {Kingdom} {Sanction} {Members} of {Russia}-{Based} {Trickbot} {Cybercrime} {Gang}},
	url = {https://home.treasury.gov/news/press-releases/jy1256},
	language = {en},
	urldate = {2023-02-09},
	journal = {U.S. Department of the Treasury},
	month = jan,
	year = {2023},
}

@misc{noauthor_conti_nodate-8,
	title = {Conti – {Rewards} {For} {Justice}},
	url = {https://rewardsforjustice.net/rewards/conti/},
	language = {en-US},
	urldate = {2022-08-19},
}


\end{document}